\documentclass[cits]{PoS}

\title{Supernovae type Ia: non-standard candles of the Universe}

\ShortTitle{SNe Ia: non-standard candles}

\author{{A. I. Bogomazov}\\%\thanks{A footnote may follow.}\\
        Sternberg Astronomical Institute, Lomonosov Moscow State University \\
        E-mail: \email{bogomazov@sai.msu.ru}}

\author{A. V. Tutukov \\
        Institute of astronomy, Russian Academy of Sciences \\
        E-mail: \email{atutukov@inasan.ru}}

\abstract{We analyze the influence of the evolution of light
absorbtion by grey dust in SNe Ia host galaxies and the influence
of the evolution of average total mass of coalescing double
carbon-oxygen white dwarfs (progenitors of SNe Ia) under the
influence of gravitational radiation on the interpretation of
Hubble diagrams of SNe Ia. Significant increase in the average
energy of SNe Ia due to increase in the total mass of merging
dwarfs can be observed at red shift z> 2. The observed dependence
of the distance modulus from the red shift in observations of SNe
Ia can be explained not only by the assumption about accelerated
expansion of the Universe, but also by the evolution of the
absorbtion of light by grey dust in various types of host galaxies
of SNe Ia, by the effects of observational selection and by the
decrease in the average mass of coalescing degenerate dwarfs.}

\FullConference{25th Texas Symposium on Relativistic Astrophysics - TEXAS 2010\\
        December 06-10, 2010\\
        Heidelberg, Germany}

\begin{document}

In the paper \cite {bogomazov2009}\footnote{In this paper we
presented the results of population synthesis made using the
``Scenario Machine'' (computer code designed for studies of the
evolution of close binary stars, a description of the code can be
found in \cite{lipunov2009}.} we showed that the average total
mass of double carbon-oxygen (CO) white dwarfs (WD) coalescing
under the influence of gravitational waves should evolve with
time. Also we calculated the frequency of mergings of CO WDs and
constructed mass distribution of merging CO WDs. See Figure 1 for
the mass distribution of CO WDs and Figure 2 for the evolution of
average total mass. Such studies are important, because at the end
of the last century two groups of SNe Ia observers concluded that
the Universe expands with acceleration
\cite{riess1998,perlmutter1999}. Consequently, the understanding
of physics SN Ia (which determines the possible evolution of their
brightness over time) is necessary to understand the fundamental
properties of the Universe. Also we consider here the influence of
the evolution of the absorbtion of light by the grey dust in host
galaxies of SNe Ia. The absorbtion of light by such dust can lead
to the same visible effect of darkening of distant SNe Ia as
expected from the accelerated expansion of the Universe.

Currently, the most popular explanation for SN Ia is mergers of
two carbon-oxygen white dwarfs (CO WDs) under the action of
gravitational radiation, provided the total mass of the merging
dwarfs exceeds the Chandrasekhar limit. This scenario for the
formation of SN Ia was suggested in the early 1980s
\cite{tutukov1984,iben1984}. Apart from mergers of two CO dwarfs,
possible origins include the thermonuclear explosion of a WD in a
semi-detached system during the accretion of matter from its
companion, when the dwarf's mass exceeds the Chandrasekhar limit,
and a thermonuclear explosion of a white dwarf with a helium donor
star companion; under certain conditions, the mass of the WD may
be lower than the Chandrasekhar limit (see, for example,
\cite{ruiter2010}). Here, however, we consider the basic scenario
for the formation of SN Ia to be the merging of two CO dwarfs
whose total mass exceeds the Chandrasekhar limit
\cite{yungelson2010}.

Brightness of SNe Ia can be weakened by the absorbtion in the
material around the precursor of a supernova, as well as by the
absorbtion by the dust in the host galaxy of the supernova. The
most important characteristics of the dust in the host galaxies of
supernovae can change over time: the total mass of the dust,
distribution of the dust in the galaxy and the composition of the
dust. Absorbtion by the interstellar grey dust can lead to
overestimated photometric distances in comparison with the
universe without the dust. Absorbtion by the intergalactic grey
dust was suggested as an explanation of distant SNe Ia darkening
instead the acceleration of the expansion of the Universe in paper
\cite{aguirre1999a} just after classical works
\cite{riess1998,perlmutter1999}.

\begin{figure}
\center\includegraphics[width=.75\textwidth]{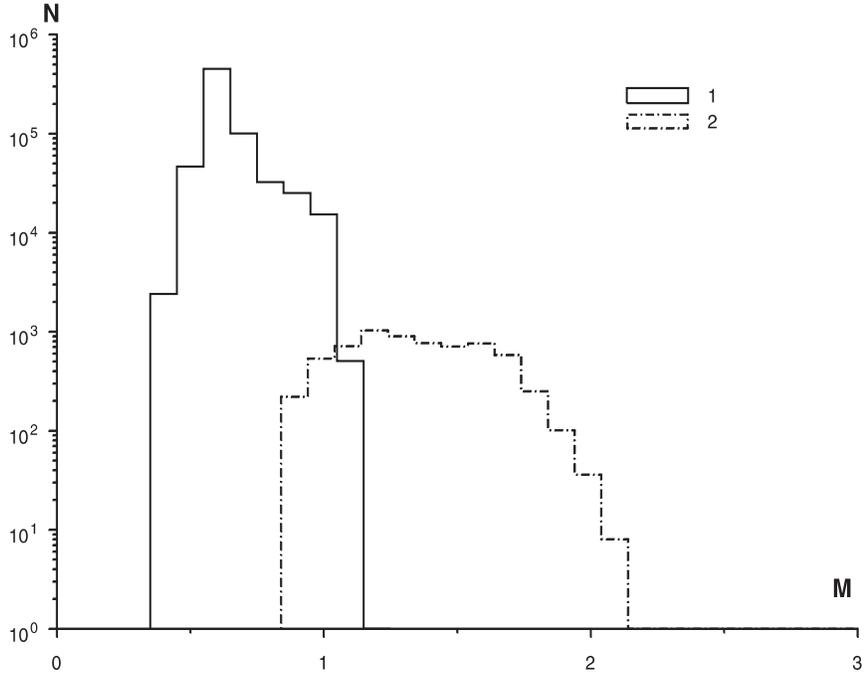}
\caption{The mass distribution for carbon-oxygen WDs that have not
undergone mergers (solid curve) and the total mass distribution
for merging WDs (dash-dotted curve). The horizontal axis plots the
mass of the WD in $M_{\odot}$ (the total mass of the two WDs in
the case of merging WDs), and the vertical axis the number of WDs
(and of mergers) derived from the calculations.} \label{f1}
\end{figure}

In our work we study the absorbtion of light by interstellar dust.
To estimate the absorbtion of light of SNe Ia in disk galaxies we
use a model of the evolution of galaxies from papers
\cite{firmani1992}-\cite{kabanov2011}. The mass of the simulated
galaxy was taken to be equal to $2\cdot 10^{11} M_{\odot}$, and
the radius $\sim 10^4$ pc. Star formation rate is estimated from
the condition of complete ionization of the gaseous disk of the
galaxy by the young massive stars. The thickness of the gaseous
disk is estimated from the condition of equality of the energy of
the SNe explosions converted into the turbulent motion of the gas
in the galaxy and by the energy dissipation rate of the turbulent
motion of the gas. Frequency of SNe explosions determines the star
formation rate in the simulated galaxy, and the dissipation
turbulent motion of the interstellar gas is determined by the
collisions of the interstellar gas clouds between each other. At
the same time it is essential that extinction in elliptical
galaxies is negligible in comparison to disk galaxies.

Is is important to note that we take into consideration so-called
``grey'' dust. Light absorbtion by such dust does not depend on
the light wavelength. This kind of dust has already been
discovered in our own Milky Way Galaxy \cite{gorbikov2010}. Small
dust particles can be taken into account due to the light
reddening at least in principle. Also it is important to take into
account that modern simulations of the evolution of the optical
depth of the galaxy are not completely defined and it is not
possible to estimate the optical depth with precision better than
by a factor of 2 \cite{firmani1992,wiebe1998}. Light absorbtion by
the grey dust does not depend on the SNe Ia model. We suppose that
the abundance of the grey dust in the host galaxies of SNe Ia is
proportional to the abundance of all kinds of dust in these
galaxies.

We assume that the average absolute magnitude of type Ia
supernovae depends on the average total mass of merging CO WDs as
follows: $M=C - 2.5\cdot \log M_{\Sigma}$, here $M$ is the average
absolute magnitude of SNe Ia, $M_{\Sigma}$ is the average total
mass of merging CO WDs, $C$ is a constant value. In Figures 3 and
4 we present the dependence of value $\Delta (m-M)$ from the red
shift $z$: $\Delta (m-M)=(m_1-C)-(m_2-C)=m_1-m_2$, here $m$ and
$M$ are visible and absolute magnitudes correspondingly, $m_1$ is
the visible magnitude, evolution of the average total mass of
merging CO WDs is taken into account, $m_2$ is the visible
magnitude calculated in supposition that SNe Ia are standard
candles with absolute magnitude $C$.

\begin{figure}
\center\includegraphics[width=.75\textwidth]{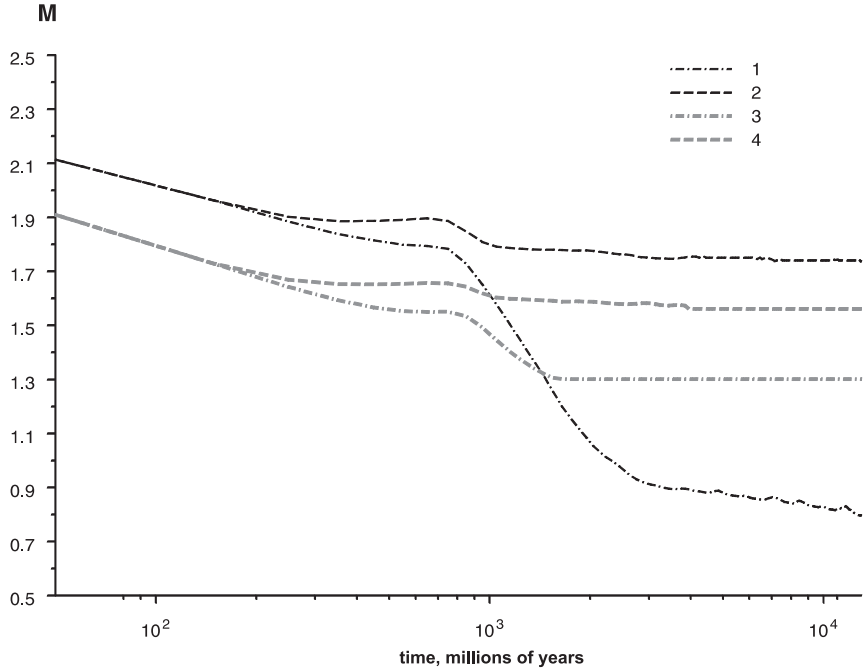}
\caption{The average mass of merging WDs as a function of time
from the onset of the star formation (in solar masses). The
numbers at the top mark the curves corresponding to mergers of 1
all WDs, 2 WDs of all possible chemical compositions, but taking
into account only mergers in which the total mass of the WDs
exceeds the Chandrasekhar limit, 3 carbon-oxygen WDs, and 4
carbon-oxygen WDs with the total mass exceeding the Chandrasekhar
limit. The star formation history similar was taken from
\cite{kurbatov2005}.} \label{f2}
\end{figure}

According to our calculations (see Figures 3 and 4) the average
energy of SNe Ia should increase with red shift at $z>2$ and such
increase should become significant at red shift $z \gtrsim 8$,
because the average total mass of merging dwarfs was greater
during early stages of evolution of the Universe. There are no
known supernovae at so high red shift. Therefore this conclusion
can not affect the result of classical papers
\cite{riess1998,perlmutter1999}. At the same it is important to
take into account that in our population synthesis models we study
the evolution of a population of stellar objects (any galaxy),
which was born at a definite moment of time (at red shift $z^*=10$
in this work), then this population evolves.  If there is a
significant burst of star formation after $z^*$ we should compare
our results with the period of time elapsed from this star
formation burst. Also independently from the population age a
dispersion of SNe Ia parameters should be due to the fact that
total mass of merging dwarfs can vary approximately by a factor of
1.5 (see Figure 1).

\begin{figure}
\center\includegraphics[width=.75\textwidth]{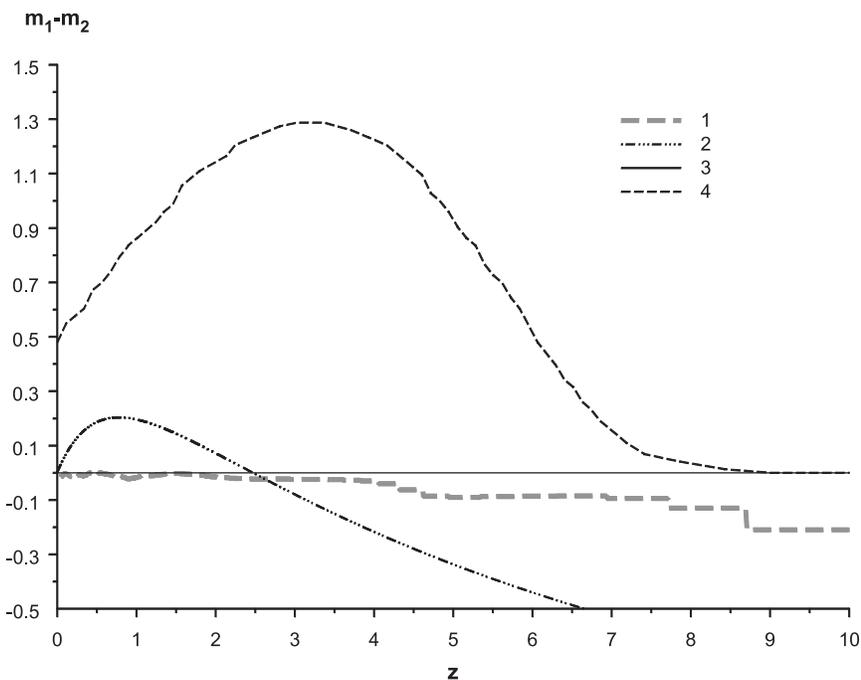}
\caption{In this figure we plot the difference $\Delta (m-M)$
between dependencies of the SNe Ia magnitude on the red shift of
SNe Ia with the effects under consideration (we consider the
evolution of absorbtion of light by dust and the evolution of
average total mass of merging CO WD) and the same dependencies
calculated under the assumption that SNe Ia are standard candles
and the absorbtion of light by dust does not evolve. The curve 1
is calculated under the assumption that there is no absorbtion and
the absolute magnitude of SNe Ia evolves from its maximum value in
the beginning of the evolution to a constant value approximately
after one billion years of evolution of stars in galaxies, it
corresponds to the curve 4 in fig. 2, we use the next set of
cosmological parameters for this curve: $\Omega_{\lambda}=0$,
$\Omega_{M}=0.2$, $\Omega_{tot}=0.2$, $H_{0}=70$ km s$^{-1}$
mpc$^{-1}$, $z^{*}=10$. The curve 2 is calculated under the
assumption that absorbtion does not play any role, the absolute
stellar magnitude of SNe Ia does not evolve and
$\Omega_{\lambda}=0.7$, $\Omega_{M}=0.3$, $\Omega_{tot}=1$,
$H_{0}=70$ km s$^{-1}$ mpc$^{-1}$. The curve 3 corresponds to the
model without evolutionary effects and $\Omega_{\lambda}=0$,
$\Omega_{M}=0.2$, $\Omega_{tot}=0.2$, $H_{0}=70$ km s$^{-1}$
mpc$^{-1}$. The curve 4 describes evolution of absorbtion of light
by dust, this dependence was taken from \cite{kabanov2011}
(dependence of absorbtion $A_v$ on red shift $z$, mass-radius
relation is $M\sim R^2$).} \label{f3}
\end{figure}

The evolution of the absorbtion of light by the interstellar dust
according to the model under consideration in this work can lead
to the much higher decrease of brightness of distant SNe in
comparison with the assumption about the dark energy existence in
the Universe. At the same time the curve 4 in Figures 3 and 4
describes the absorbtion by all kinds of dust. If we change
internal parameters of the dust evolution model and set up
required abundance of grey dust we can make (in principle)
practically ideal coincidence between curves 2 and 4 in a very
wide range of red shift. Nevertheless such method is under
critique, because fine tuning of the model parameters is an
unattractive way to explain physics \cite{riess2007}.

It was shown that SNe Ia are $\approx 0^m.1$ brighter in quiet
galaxies than in galaxies with active star formation
\cite{lampeitl2010}. The dependence between the SNe Ia brightness
and the type of SNe Ia host galaxies also was mentioned in the
paper \cite{sullivan2010}.

\begin{figure}
\center\includegraphics[width=.75\textwidth]{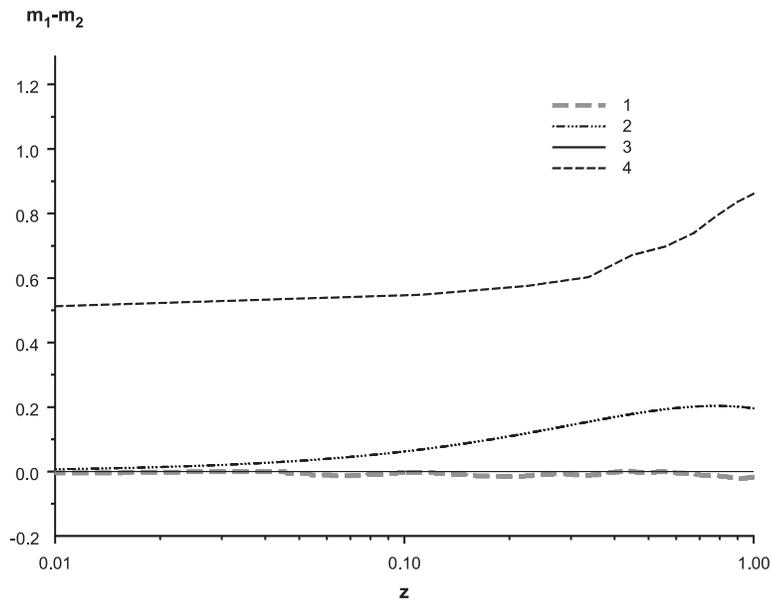}
\caption{The same as Figure 3 in the red shift interval
z=$0.01\div1$.} \label{f4}
\end{figure}

We offer the following interpretation of our results. As we can
see from the figure 3, the growth of the dust absorbtion in disk
galaxies takes place at almost the same rate as the additional
darkening of the supernovae light grows with the redshift due to
the assumption of accelerated expansion of the Universe.
Consequently, it is possible to hypothesize that the accelerated
darkening of the supernova (in average, which is important) is the
result of the evolution of the absorbtion of radiation by the grey
dust, the effect of which is not taken into account yet. However,
with further increase of the red shift the absorbtion also
increases. This means that SNe fade more and more in comparison
with the model without dust. There are a lot of distant
supernovae, which belong to the slow down curve rather than to the
curve of accelerated expansion of Universe (see, for example,
figure 10 in paper \cite{kowalski2008}). This fact is an argument
against the explanation of the accelerated darkening of SNe Ia by
the evolution of the absorbtion of light by grey dust. But this
argument is valid only if all host galaxies of SNe Ia are disk
galaxies with active star formation. This is not so: for example,
host galaxy of SN 1997 ff (this is one of the most distant SNe Ia)
is a quiet elliptical galaxy \cite{riess2001}.

As we already mentioned, the absorbtion in elliptical galaxies in
comparison with disk galaxies is negligible and it does not evolve
during long period of time comparable with Hubble time. We also
would like to note that according to population synthesis results
(see, for example, figure 1 in paper \cite{jorgensen1997}) the
frequency of SNe Ia in an elliptical galaxy drops approximately by
a factor of two between $1$ and $10$ billions of years from the
birth of the galaxy. This means that in past the probability of SN
Ia explosion is higher in an elliptical galaxy than in a disk
galaxy. This fact probably explains observed absence of
``accelerated'' SNe Ia with red shift $z>1$.

Thus, the use of SNe Ia as cosmological standard candles must take
into account not only characteristics of the model of supernovae,
but also the evolution of the absorbtion by dust (including the
grey dust) in the host galaxies of SNe Ia. It is better to use SNe
Ia in galaxies without star formation as distance indicators. SNe
Ia in galaxies with active star formation can be used as a probe
to study the evolution of the absorbtion in these galaxies.

\end{document}